\documentclass[twoside,twocolumn,english,leqno]{article}
\usepackage[]{fontenc}
\usepackage[latin1]{inputenc}
\usepackage{amsmath}
\usepackage{graphicx}
\usepackage{amssymb}

\makeatletter

\providecommand{\LyX}{L\kern-.1667em\lower.25em\hbox{Y}\kern-.125emX\@}
\newcommand{\noun}[1]{\textsc{#1}}

\newtheorem{Theorem}{Theorem}[section]

\newtheorem{crly}[Theorem]{Corollary}
\def\qed{\hfill $\Box$ \vskip 3pt}

\usepackage{epsfig}
\usepackage{ltexpprt}
\usepackage{latexsym}
\makeatletter
\let\ps@plain=\ps@empty
\makeatother

\usepackage{babel}
\makeatother
\begin{document}

\title{The Wake Up and Report Problem is Time-Equivalent to the Firing Squad
Synchronization Problem\date{}}

\author{Darin Goldstein%
\footnote{Computer Science Department, Cal State Long Beach, daring@cecs.csulb.edu%
} and Nick Meyer%
\footnote{Mathematics Department, U.C. Berkeley, meyer@math.berkeley.edu%
}}

\maketitle
\begin{abstract}
We consider several problems relating to strongly-connected \emph{directed}
networks of identical finite-state processors that work synchronously
in discrete time steps. The conceptually simplest of these problems
is the Wake Up and Report Problem; this is the problem of having a
unique {}``root'' processor send a signal to all other processors
in the network and then enter a special {}``done'' state only when
all other processors have received the signal. The most difficult
of the problems we consider is the classic Firing Squad Synchronization
Problem; this is the much-studied problem of achieving macro-synchronization
in a network given micro-synchronization. We show via a complex algorithmic
application of the {}``snake'' data structure first introduced in
Even, Litman, and Winkler\cite{ELW} that these two problems in particular
are asymptotically time-equivalent up to a constant factor. This result
leads immediately to the inclusion of several other related problems
into this new asymptotic time-class. 
\end{abstract}

\section{Introduction}

In this paper, we study a number of fundamental protocols that run
on \emph{directed} networks of identical synchronous finite-state
automata. Two of the most important are the Wake Up and Report Problem
(or WURP, also known equivalently in the literature as the {}``Broadcast
With Echo'' Problem%
\footnote{The authors had a touch choice to make between WURP and BWEP. We chose
WURP. The reader is welcome to weigh in on his acronym of choice.%
}), conceptually the simplest problem to solve, and the Firing Squad
Synchronization Problem (or FSSP), probably the most difficult. 

Generally speaking, the bulk of researchers studying theoretical protocols
on such networks tend to focus on the seemingly more difficult and
intricate FSSP which has a forty-year history of research behind it.
The WURP, though obviously important, seems to receive less attention
because of its apparent simplicity. The recent research trend (e.g.
\cite{ELW}, \cite{OW}) seems to be for a group to discover a clever
solution to the FSSP and then notice that the solution also solves
a number of other problems including the WURP. Our main result is
the introduction of a new time-class of asymptotically equivalent
network problems which includes both the WURP and the FSSP. This result
makes it possible for researchers to reverse the trend. Instead of
working on the {}``difficult'' problems and noting that the solutions
solve the {}``easy'' ones, researchers can now focus their efforts
on {}``easy'' problems and note via the result in this paper that
asymptotically those solutions also solve the {}``difficult'' ones.
One can, of course, look at this benefit from the opposite direction:
a \emph{lower} asymptotic time bound for one of the {}``difficult''
problems also applies to the {}``easy'' ones.

\subsection{The Network Model. }

We consider strongly-connected \emph{directed} networks of identical
synchronous finite-state automata with in- and out-degree bounded
by a constant. These automata are meant to model very small, very
fast processors. The network itself has unknown topology and potentially
unbounded size $N$. Throughout this paper, we use the term {}``-port''
to refer to one of a number of unidirectional conduits through which
constant-size messages may pass from one processor to another. An
\emph{in}-\emph{port} to a processor will allow messages to flow unidirectionally
in towards the processor. \emph{Out}-\emph{port} is defined similarly.
We assume throughout that the number of in-ports and out-ports for
each processor is uniformly bounded by a network constant $\delta $.
The network is formed by connecting out-ports from automata to the
in-ports of other automata with wires. Not all in-ports or out-ports
of a given automaton need necessarily be connected to other automata.
Note that even though the communication links are unidirectional,
a pair of processors is allowed to be connected with two communication
links, one in either direction, simulating a bidirectional link. 

We assume that each processor in the network is initially in a special
{}``quiescent'' state, in which, at each time-step, the processor
sends a {}``blank'' character through all of its out-ports. A processor
remains in the quiescent state until a non-blank character is received
by one of its in-ports. 

The network has a global clock, the pulses between which each processor
performs its computations. Processors synchronously, within a single
global clock pulse, perform the following actions in order: read in
the inputs from each of their in-ports, process their individual state
changes, and prepare and broadcast their outputs. 

Our network structure is specifically designed to model the practical
situation of many small and fast processors with only the capacity
for reliable one-way communication. Aside from being intrinsically
mathematically interesting in its own right, this particular model
applies to many practical situations as well. Unidirectional communication
commonly occurs with one-way radio networks (e.g. the GPS satellites
currently in orbit about Earth, encrypted military networks, etc.),
VLSI design, and bidirectional networks in which in-ports or out-ports
are allowed to undergo shutdown failures. 

The reason for modeling the processors by identical finite-state automata
is simple. In practice, many network protocols are expected to run
extremely fast. One particular reason for this is that the network
topology or size might change if the protocol takes too long, thereby
potentially rendering the computation obsolete or incorrect. (For
example, if the network is attempting to synchronize itself and a
processor is randomly added to the topology of the network in the
middle of the computation, it is likely to throw off the timing of
the synchronization.) It is therefore assumed that the processors
involved will not have time to access a large memory cache. Commonly,
a memory access can take orders of magnitude longer than a simple
state-change processor calculation. The current technological trend
is to merge the memory functions that one generally associates with
a computing machine into the processor itself (for further discussion,
see \cite{awerbuch94memoryefficient}, \cite{mayer92selfstabilizing}.)
Hence, we model the processors by identical finite-state automata. 

The protocols described below are presumed to begin when a certain
processor is signaled by some outside source. We call this processor
the \emph{root}, and assume that every processor knows whether or
not it is the root. (It is possible to do this by specifying a {}``root
subset'' of the state space of the automata that the root is uniquely
allowed to use. The root's state must always lie in this special root
subset, and no non-root processor's state can ever be in the root
subset.) A protocol ends when the root enters a special terminal state
indicating that the computation has successfully completed. In our
computational model, we calculate the time-complexity of a protocol
in terms of the total number of global time steps between initiation
and termination. Of course, the aim is to minimize this time-complexity. 

Finally, for ease of exposition, we define several statistics associated
with a given network: 

\begin{Definition}\label{dfn:gamma}

Let \emph{distance} in the network be defined in the obvious way;
i.e., define \(d(A,A')\), the distance from processor \(A\) to processor
\(A'\), to be the length (in number of edges) of a minimal-length directed
path from processor \(A\) to processor \(A'\). (Note that we are on
a directed network, so \(d(A,A')\) might not equal \(d(A',A)\).)

We define \(\Gamma := \max _{\{\mbox {processors\, }A\}}d(\mbox {root},A)\),
\(\Gamma ':= \max _{\{\mbox {processors\, }A\}}d(A,\mbox {root})\), and
\(D := \max _{\{\mbox {all processors\, }A,B\}}d(A,B)\).

\end{Definition}

Note that if we start a breadth-first search at the root, then $\Gamma $
is the length of the longest branch in the resulting BFS tree. 

We point out that none of the finite-state processors comprising our
network may be able to store $\Gamma ,\Gamma ',$ or $D$, as they
may be arbitrarily large.

\section{Description and Brief History of Prior Research}

Research on networks of identical finite-state automata has been going
on for decades, and the body of literature on the subject is enormous.
Aside from the two problems presented below, the literature deals
with issues such as leader election, mutual exclusion, and network
traversal. See \cite{afek94distributed}, \cite{angluin80local},
\cite{codenotti95symmetry}, and \cite{singh91efficient} for a \emph{small}
sampling. The remainder of this section will be devoted to a description
and brief history of the prior work relating to the two main problems
we address in this paper, the Wake Up and Report Problem and the Firing
Squad Synchronization Problem.

\subsection{The Wake Up and Report Problem. }

Conceptually, the Wake Up and Report Problem (or WURP) is the simpler
of the two problems we will consider in this section. We begin with
a network of quiescent processors. Some outside agent nudges the root
processor to commence the Wake Up and Report Protocol. The root may
terminate the protocol after every processor in the network has been
nudged out of the quiescent state. The objective of the protocol is
therefore to inform the root in the fastest possible time that every
processor has {}``awakened'' from its quiescent state. This is equivalent
to the Broadcast and Echo demand that the root be assured that every
processor in the network has received a unique signal sent out upon
protocol initiation. 

As mentioned previously, this problem has not been studied to the
same extent as the Firing Squad Synchronization Problem below. The
best solution currently known is the $O(ND)$ protocol outlined in
Ostrovsky and Wilkerson's 1995 paper\cite{OW}. Aside from the references
to the Wake Up and Report problem in Even, Litman, and Winkler\cite{ELW}
and Ostrovsky and Wilkerson\cite{OW} (both papers mainly dealt with
the more difficult FSSP), references to the Wake Up and Report Problem
can be found with variations in the network and processor assumptions
under the guise of the Broadcast and Echo Problem in Afek and Gafni\cite{afek94distributed},
Propagation Information with Feedback in Segall\cite{segall83distributed},
and Echo protocols in Chang\cite{chang82echo}. We should take care
to mention that in some of these references, the presented protocols
were more concerned with message complexity than time complexity.
This paper will concentrate exclusively on the time complexity aspect
of the problem. 

The only fact about the WURP that we will need is the following intuitively
obvious lemma, which states that in order for a WURP protocol to terminate
correctly on every network, the root must have had time to both send
out and receive a message from every other processor in the network. 

\begin{lemma}

\label{lemma:LBL}The running time of any WURP protocol must be greater
than \(\max (\Gamma ,\Gamma ')\).

\end{lemma}

\emph{Proof%
\footnote{This proof can be skipped without much loss to the reader; we bother
to prove this intuitively obvious lemma rigorously because it will
be crucial to the proof of our protocol's correctness.%
}.}  Fix some Wake Up and Report Protocol, fix a network, and let $t$
be the running time of the protocol on this particular network. Assume
$t\leq \Gamma $. Fix a processor $G$ such that $d(\mbox {root},G)=\Gamma $.
Now, construct a new network from the old, replacing $G$ with a length-2
chain of processors (call them $G_{1}$ and $G_{2}$), such that the
inputs of $G$ enter the chain at $G_{1}$ and the outputs of $G$
exit the chain at $G_{2}$. 

\begin{figure}
\begin{center}\includegraphics[  scale=0.45]{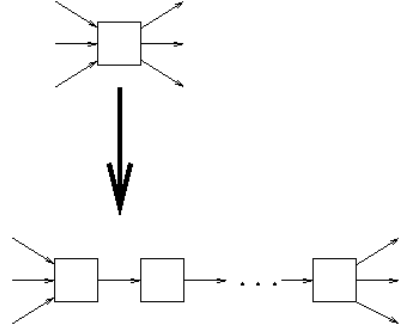}\end{center}

\caption{The replacement of a processor with a sequence of processors\label{figure:links}}
\end{figure}

Now, the computational transcript of the root in our new network is
identical to that of our old network, up through time $\Gamma $;
hence, our protocol should once again terminate at time $t$. But
this leads to a contradiction -- since $d(\mbox {root},G_{2})=\Gamma +1$,
it is impossible for processor $G_{2}$ to be awakened by time $t$. 

Now assume $t\leq \Gamma '$. Fix a processor $G'$ such that $d(G',\mbox {root})=\Gamma '$;
define $t_{0}$ to be $d(\mbox {root},G')$. Let $D$ be the diameter
of the network. Construct a new network from the old, replacing $G'$
with a chain of $D+1$ processors (call them $G_{1}'$ through $G_{D+1}'$,
such that the inputs of $G'$ enter the chain at $G_{1}'$ and the
outputs of $G'$ exit the chain at $G_{D+1}'$. (See Figure \ref{figure:links}.)
Clearly, the computational transcript of the root is identical to
that with our old network, up through time $t_{0}+\Gamma '-1$, which
is certainly greater than or equal to $\Gamma '$; hence, as before,
the protocol will terminate at time $\Gamma '$. But this leads to
a contradiction -- since $d(\mbox {root},G_{D+1}')=t_{0}+D>\Gamma '$,
it is impossible for processor $G_{D+1}'$ to be awakened by time
$\Gamma '$. \qed

This lemma leads to an interesting corollary that, once again, is
obviously true on an intuitive level but worth proving explicitly.
This Corollary makes explicit the fact that either $\Gamma $ or $\Gamma '$
must be at least $\frac{D}{2}$. (Note that it is easily possible
to construct families of networks where $\Gamma =O(\log D)$ or $\Gamma '=O(\log D)$.
Otherwise, the little substance in this Corollary would disappear
entirely.) 

\begin{crly}\label{crly:LBC}

Any protocol that solves the Wake Up and Report Problem must take
time at least \(\frac{D}{2}\), where \(D\) is the diameter of the network.
Hence, any such protocol has time-complexity \(\Omega (D)\).

\end{crly}

\emph{Proof.} Fix a WURP protocol and a network. Assume the protocol
has running time $t$ on the given network; fix processors $A$ and
$B$ in our network such that $d(A,B)=D$. Then, from Lemma \ref{lemma:LBL},
we have $D\leq d(A,\mbox {root})+d(\mbox {root},B)\leq \Gamma +\Gamma '\leq 2t$,
from which the result follows. \qed

\subsection{The Firing Squad Synchronization Problem.\label{subsection:FSSPhistory} }

Again, we have a network as described previously. An outside agent
nudges the root out of its quiescence. The goal of a Firing Squad
Synchronization Protocol is to get every processor in the network
to \emph{simultaneously} enter the the same special FIRING state.
Termination of the protocol occurs when the root (and hence the rest
of the network) has FIRED. This protocol is useful for global calculations
which require all processors to begin computation simultaneously.
What makes this problem fiendishly difficult is the finite-state property
of the automata: no single processor has the memory capacity to determine
the size or topology of a large enough portion of the network. In
general, processors do not even necessarily have the ability to assign
themselves unique names if the network is large enough (as it may
well be.) 

The Firing Squad Synchronization Problem (or FSSP) has a rich history;
solutions to various subproblems have been discovered over a period
of decades. J. Mazoyer provides an overview of the problem (up to
1986, at least) in addition to some of its history in \cite{Maz}.
We summarize the history here. J. Myhill introduced the problem in
1957, though the first published reference is \cite{Moore} from 1962.
J. McCarthy and M. Minsky first solved the problem for the bidirectional
line in \cite{McMin}. Later, the problem was considered for directed
networks of automata. Honda and Nishitani\cite{HN} and Kobayashi\cite{Kob}
solved the FSSP for specific subtypes of networks, the directed ring
and the {}``ring-of-trees'' -- that is, networks which include a
loop, containing the root, whose length is at least as great as the
maximum distance from the root to any processor. Even, Litman, and
Winkler\cite{ELW} used this, and their invention of network-traversing
{}``snakes'' (see below) to create a protocol that would fire any
strongly-connected directed network in $O(N^{2})$ time. Ostrovsky
and Wilkerson\cite{OW} were able to improve this to $O(ND)$ in 1995,
which remains the best today.

\section{Data Structures}

Space does not permit us to expand upon the fascinating history outlined
in Section \ref{subsection:FSSPhistory}; in particular, we cannot
present the various solutions to the FSSP that have been found over
the years. Our protocol does make use of several data structures discovered
and refined by those mentioned above; we review those structures here. 

We first describe how to implement data structures of differing {}``speeds''.
Then we briefly discuss tokens, various types of snakes, and the loops
that snakes generate and mark. Finally, we define the {}``ring-of-trees''
network structure, present a way to impose this structure on the network
(given a sufficiently long marked loop), and discuss how this will
enable us to fire the network quickly.

\subsection{Speed. }

The protocol about to be presented makes use of several computational
constructs (to be described below), each of which is assigned a certain
characteristic {}``speed.''%
\footnote{The concept of speed was introduced by McCarthy and Minsky \cite{McMin}
in a paper in which they presented the first published solution to
the FSSP on a bidirectional line of processors.%
} This is \emph{not} to say that certain messages move faster through
the network than others. All computations and outputs are strictly
synchronous with respect to the global network clock. 

In the protocol that follows, the speeds that we utilize are speed-1,
and speed-4. The method by which we implement a speed is as follows:
A speed-1 construct will enter a processor. It will then remain there
for 4 global clock ticks. At the 4th clock tick, it will proceed along
its designated path. A speed-4 construct, on the other hand, will
wait only for one tick. Thus, in reality, the implementation specifies
that a speed-1 construct moves 4 times slower than a speed-4 construct.

\subsection{Tokens.\label{subsection:tokens} }

Tokens are the simplest data structure possible on networks of finite-state
machine. They should be thought of as markers that can be passed from
one processor to another via the edges of the network. The token concept
has been in use since the first solution to the bidirectional line
\cite{McMin}. The general token behaviors outlined below have also
been utilized in other papers (e.g. \cite{ELW} and \cite{OW}), though
our definition of {}``breadth-first'' and {}``loop'' tokens below
are original to the best of our knowledge. 

We employ two main varieties of tokens. \emph{Breadth-first tokens}
can be thought of as moving within a {}``breadth-first-search tree''
in the following sense: We arrange it so that each relevant processor
in the network has a {}``parent'' marker associated with one of
its in-ports%
\footnote{The method by which various breadth-first-search trees are constructed
by snakes, as well as how each processor designates its \char`\"{}parent\char`\"{}
in-port, is discussed in Section \ref{section:algorithm}.%
}. We then declare that a breadth-first token will only be accepted
by a given processor when either (a) the processor creates the token,
in which case the processor will not have a parent in-port, or (b)
the token comes through the processor's parent in-port. If a breadth-first
token is received through a non-parent in-port or by a quiescent processor,
it is ignored. Breadth-first tokens are passed out of every out-port;
thus, breadth-first tokens multiply in number as time goes on (as
long as they stay within the confines of the breadth-first-search
tree.) In summary, if a breadth-first token is created at the root
of its associated breadth-first tree, then $t$ time steps later there
will be a token at each processor that is a distance of $t$ from
the root, and none elsewhere. (If the tree has length less than $t$,
of course, there will be no tokens anywhere.) 

\emph{Loop tokens} travel along a specified marked loop within the
network. (How loops get marked is described in Section \ref{subsection:snakes}.)
A processor on the loop that receives a loop token simply passes it
to the next processor on the loop. In this paper, the root (which
will be on every loop we mark) will create every loop token we use;
$t$ time steps after its creation, therefore, a loop token will be
$t$ processors away from the root, along the marked loop. When any
loop token reaches the root again, it is absorbed (i.e., not sent
around again). 

Note that tokens can only carry along with them a constant (very small)
amount of information since they are only of constant size. The next
data structure takes care of this problem.

\subsection{Snakes.\label{subsection:snakes} }

The concept of a data-carrying \emph{snake} was invented by Even,
Litman, and Winkler in \cite{ELW}. Snakes are the solution to the
problem of the limited data-carrying capabilities of tokens. A snake
is capable of carrying an arbitrarily large amount of data, but for
this reason, it must reside in a collection of adjacent processors
rather than a single processor. 

A {}``snake'' is a string -- which may be arbitrarily long -- made
from an alphabet of $2\delta +1$ characters, namely $\delta $ head
characters, $\delta $ body characters, and a unique tail character.
(Recall that $\delta $ is a fixed constant of the network.) The characters
comprising the string are stored in adjacent processors, one character
per processor. These characters encode a path by specifying a series
of out-ports. (Note that a token could never do such a thing, since
a path in the network can grow arbitrarily long.) 

We require two main snake types, which we call growing and dying%
\footnote{Even, Litman, and Winkler in \cite{ELW} refer to growing and dying
snakes as adders and rattlers, respectively.%
}. \emph{Growing snakes} are used to generate encoded paths of the
network, and \emph{dying snakes} are used to mark encoded paths. Our
protocol requires two kinds of each of the two snake types; specifically,
we will need out-growing, in-growing, out-dying, and in-dying snakes.
{}``Out'' and {}``in'' are meant as a mnemonic; out-snakes are
generated at the root and proceed outward from it, while in-snakes
are generated elsewhere and trigger some action when they reach the
root. Out-growing, in-growing, out-dying, and in-dying snakes will
be referred to as OG-snakes, IG-snakes, OD-snakes, and ID-snakes in
what follows. 

Each of the four kinds of snake gets its own alphabet of characters
to describe it; this allows processors to determine with which kind
of snake they are dealing. We will spend a section on each type, elucidating
its respective behavior. First, we need to go over some general rules
common to all snake types.

\subsubsection{General Snake-handling Rules. }

\begin{itemize}
\item All snakes are speed-1. 
\item Snakes of different types do not interact. A processor can handle
different snake types simultaneously without getting confused because
snake types are distinguished by their alphabets. Note that this does
not impose arduous memory constraints upon the processors (which are
finite-state machines) since there is only a constant number of snake
types. 
\end{itemize}

\subsubsection{Growing Snakes. }

Growing snakes function as information generators. We define the \emph{initiator}
to be the processor from which the growing snakes first emanate. The
\emph{terminator} is defined to be the processor that the snakes are
attempting to reach. Growing snakes grow in a breadth-first manner;
the first growing snake to reach the terminator processor will have
encoded within its body a minimal-length path from the initiator to
the terminator. Upon reaching the terminator, a growing snake head
might then initiate some further action based on the protocol and
the state of the terminator. The rules for handling out-growing snakes
are outlined below; the rules for handling in-growing snakes are identical
(just replace {}``OG'' with {}``IG''), except where noted. %
\begin{figure}
\includegraphics[  width=7in,
  height=3in,
  angle=90,
  origin=c]{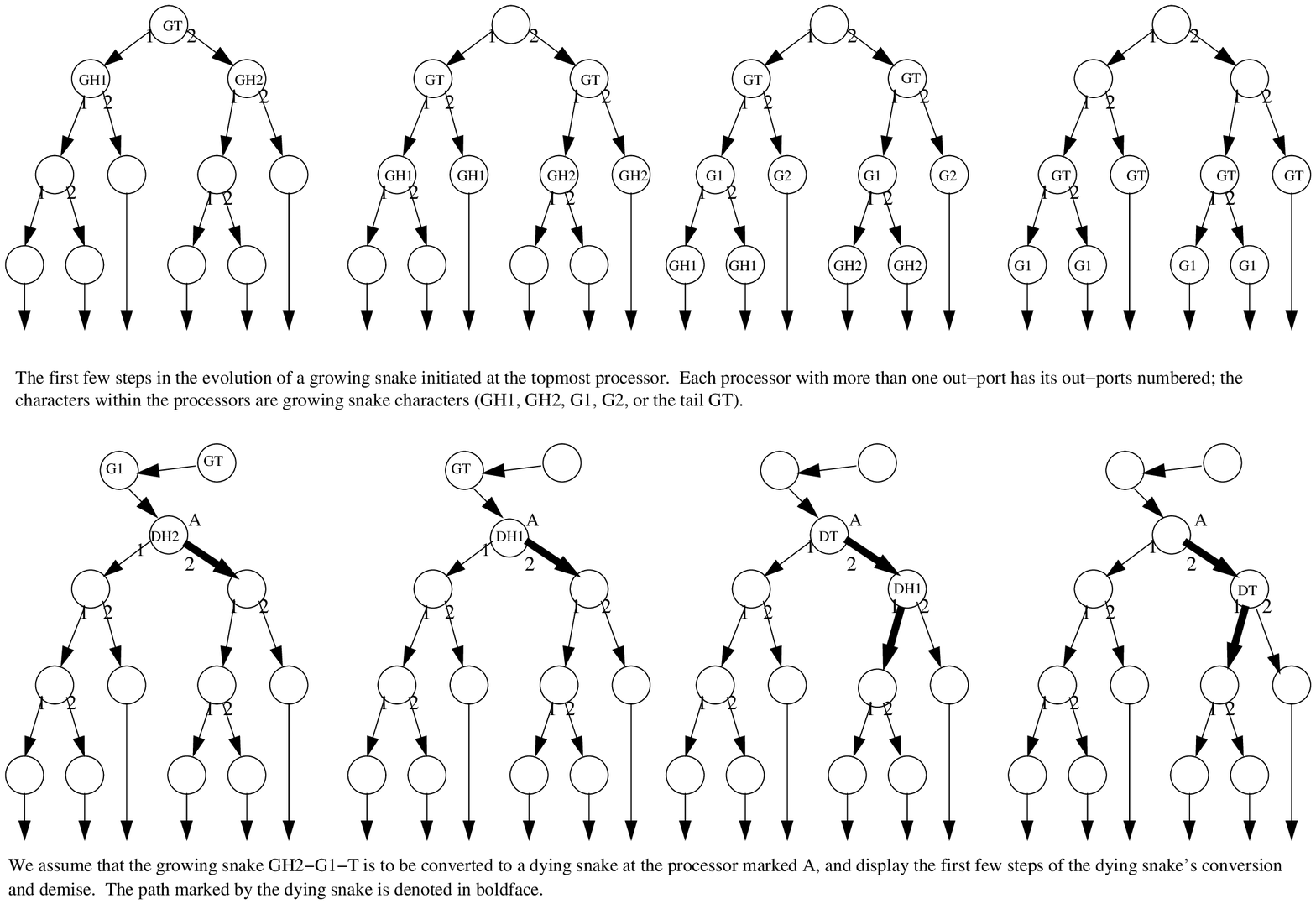}\label{figure:snakes}
\end{figure}

\begin{itemize}
\item First, the head characters of the baby OG-snakes are generated by
the initiator. This processor sends an OG-snake head character out
of every out-port during the first time step. The particular head
character to be sent will correspond to the out-port from which it
is being sent. For every $i$ between 1 and $\delta $, the growing
head snake character $OGH(i)$ will be sent through out-port $i$.
During the next time step, the initiator will send a tail character
$OGT$ through every out-port. Thus a baby snake is born. 
\item When a processor receives an out-growing snake character for the first
time, it marks itself $OG$-visited, and marks the in-port through
which the growing snake character was passed as its $OG$-parent%
\footnote{If two or more OG-snakes arrive simultaneously, the one arriving through
the lowest-numbered in-port is deemed {}``first.''%
}. (These marks will be used later by certain breadth-first tokens;
see Section \ref{subsection:tokens}.) Only this first OG-snake will
be allowed to pass through the processor; all other OG-snake characters
will be ignored. The OG-visited markings will never be removed; thus,
an OG-snake will carve out a breadth-first-search tree. IG-snakes
behave identically, except that IG-visited and IG-parent markings
are removed from the network at one stage of our protocol, thus allowing
a given processor to be visited by more than one IG-snake during its
lifetime. Until this removal occurs, however, each IG-snake will also
carve out a breadth-first-search tree. 
\item When a processor receives a non-tail OG-snake character, it simply
sends this character through all out-ports during the next time step.
Once the processor sends the character out, it need not retain it
in {}``memory.'' (In this way, the processor simply passes the head
and body through every out-port. Thus arbitrarily long paths can be
generated.) 
\item Once a processor receives the tail of an OG-snake, instead of simply
passing it through like the other body characters, the processor generates
a new body character. For all $i$ between 1 and $\delta $, it simultaneously
sends the character $OG(i)$ through out-port $i$; thus a new body
character is generated to mark the current processor's position in
the path. Only after this new character is passed along does the processor
send the tail through. 
\end{itemize}

\subsubsection{Dying Snakes.\label{subsubsection:dyingsnakes} }

Dying snakes function as path markers. After a path is generated by
the growing snakes, it is the responsibility of the dying snake to
mark the generated path so that the processors on it will know (a)
that they lie along a special path and (b) which in-port and out-port
they should use for funneling information along the path. In our protocol,
OD-snakes will be formed by converting the characters of an IG-snake
into OD-snake characters as they pass through the root; ID-snakes
will be created from OD-snakes in a manner to be described in Section
\ref{section:algorithm}. The rules for handling OD-snakes are outlined
below; the rules for handling ID-snakes are identical (just replace
{}``OD'' with {}``ID''), except where noted. 

\begin{itemize}
\item An OD-snake will mark a path generated by an OG-snake (see Section
\ref{section:algorithm}); thus, since an OG-snake carves out a breadth-first-search
tree, the path will never self-intersect. Similarly, neither will
a path that an ID-snake is to mark. However, the concatenation of
the two paths (which, in our protocol, will always be a loop beginning
and ending at the root) may self-intersect; any processor will appear
at most twice on the concatenation. We will, eventually, want to consider
the concatenation as a whole; to this end, we imbue each processor
with two {}``predecessor in-ports'' (numbered 1 and 2) and two {}``successor
out-ports'' (ditto). 
\item Whenever a processor receives the head character $ODH(i)$ of a OD-snake,
it sets predecessor in-port \#1 equal to the number of the in-port
through which it received the character, and sets successor out-port
\#1 equal to $i$. These two values indicate the two edges of the
path incident to the processor. ID-snakes work identically, except
that they use predecessor in-port \#2 and successor out-port \#2.
The head character is then discarded (not sent through any out-port). 
\item If the next OD-snake character that the processor receives through
the predecessor in-port is $OD(j)$, it gets sent through the successor
out-port as $ODH(j)$. (In other words, the next OD-snake body character
that comes through gets converted to the new head.) The processor
then passes all further OD-snake characters received through its predecessor
in-port to its successor out-port exactly as received. If the next
character happens to be a tail, then it gets sent through the successor
out-port as is%
\footnote{In our protocol, OD-snakes will be converted into ID-snakes at those
processors marked IG-start; this will provide an exception to these
rules, for at those processors all OD-snake characters are converted
into ID-snake characters instead. In addition, as might be expected,
an IG-start-marked processor will set predecessor in-port \#1 and
successor out-port \#2 appropriately as the dying snakes go through;
its other two ports will not be needed, as we will show in Section
\ref{section:algorithm} .%
}. 
\item An OD-snake only propagates along the path it is marking, and a maximum
of one will be in the network at any given time, so we need not worry
about OD-visited markings. 
\item If a processor receives only the tail of an OD-snake, it is an indication
that the processor is special in some manner depending on the specifics
of the protocol. 
\end{itemize}

\subsection{Marked loops.\label{subsection:markedloops} }

As mentioned in Section \ref{subsubsection:dyingsnakes}, we will
be using dying snakes to mark certain loops (not necessarily simple)
that begin and end at the root. We will refer to this structure repeatedly
throughout the paper, and thus make the following definition: 

\begin{Definition}

A \emph{marked loop} will be a loop marked by dying snakes in the manner
described in Section \ref{subsubsection:dyingsnakes}.  The root must be one
of the processors on the loop.  The loop may or may not be simple, but no
processor or edge can appear more than twice on it.

\end{Definition}

Each processor will have its predecessor and successor port (or, if
necessary, ports) set by the dying snakes. A processor with only predecessor
in-port \#1 set will only accept a loop token through that in-port;
it will subsequently pass the token through successor out-port \#1%
\footnote{Once again, processors marked IG-start will provide an exception to
this rule; they will accept a loop token only through predecessor
in-port \#1, but will pass it through successor out-port \#2;%
}. Similarly, a processor with only predecessor in-port \#2 set will
only accept a loop token through that in-port; it will subsequently
pass the token through successor out-port \#2. Finally, a processor
with both predecessor in-ports set will initially accept a given loop
token only through predecessor in-port \#1 (it will pass the token
through successor out-port \#1, of course); it then waits for the
token to come through predecessor in-port \#2 (at which point it passes
the token through successor out-port \#2); it then will expect the
next such loop token through predecessor in-port \#1 again. 

We will hereon refer to the predecessor in-port (resp. corresponding
successor out-port) through which a loop processor awaits a loop token
as the \emph{appropriate} predecessor in-port (resp. successor out-port).

\subsection{The Ring-of-Trees.\label{subsection:ring-of-trees} }

\begin{Definition}

A \emph{ring-of-trees} is a network consisting of a directed ring with trees 
hanging from the processors of the ring. The edges of the trees all point 
towards the leaf processors; note that, unless all of the trees are empty,
the ring-of-trees is not strongly-connected. The root processor is one of the processors on the ring, and we require that the shortest path within the structure from the root processor to any leaf processor be less than the number of processors in the ring.  This
data structure and the following discussion were introduced in \cite{HN}.

\end{Definition}

A method for firing a directed ring of processors in $O(P)$ time
(where $P$ represents the number of processors on the ring itself)
has been known for decades; we direct the reader's attention to the
references mentioned in Section \ref{subsection:FSSPhistory}. It
was shown in \cite{HN} that a ring-of-trees with $P$ processors
in its directed ring can be fired in precisely the same amount of
time; we omit the fairly simple proof. 

Now, if we can realize our network as a ring-of-trees that contains
all of the network's processors, then we'll be able to fire it. How
might this realization be done? Certainly no single finite-state processor
can hold the entire structure of the ring-of-trees. However, this
is not necessary; to pass information appropriately around the ring-of-trees,
we need only that each processor know certain of its neighbors within
the structure. Specifically, we will require that each processor on
the directed ring portion of the ring-of-trees%
\footnote{The ring, it should be noted, need not be simple, as long as no processor
is included more than a fixed constant number of times; the arguments
from \cite{HN} still work.%
} knows its predecessor in-port(s) and successor out-port(s), and that
each tree processor knows its parent in-port. This will make it easy
to transfer information within the ring-of-trees structure: each processor
on the ring can send information around it via the appropriate predecessor
and successor ports, and can send information to the trees via all
non-successor ports; each tree processor can simply send information
through all out-ports. If each processor only accepts information
through its appropriate predecessor or parent, the ring-of-trees structure
will be realized. 

So how do we find a directed ring within the network? We will use
growing and dying snakes to mark a loop into our network (see Section
\ref{subsection:markedloops} for the definition of a marked loop,
and Section \ref{section:algorithm} for the specifics of how we use
snakes to do this); this marked loop will serve as our directed ring.
We claim that, given a marked loop passing through the root with length
greater than $\Gamma $, we can realize the ring-of-trees within the
network. To do so, we use the following procedure. Each processor
on the marked loop already has a predecessor in-port (or in-ports).
The root releases a copy of a TREE-PARENT token through each of its
$\delta $ out-ports. This token acts as a cross between a breadth-first
and a loop token; upon receiving it, a processor will perform one
of the following actions, depending on its state. 

\begin{itemize}
\item If the processor already has a parent designation and is not on the
marked loop, the TREE-PARENT token is simply ignored. 
\item If the processor has no parent designation when the TREE-PARENT token
arrives, the processor designates the in-port through which the token
was received as its parent. It then passes a copy of the token through
each of its out-ports. 
\item If the processor is on the marked loop (and hence has predecessor(s)),
it checks whether the TREE-PARENT token came through its appropriate
predecessor in-port; if so, it passes a copy of the token through
its appropriate successor out-port and every non-successor out-port.
If the token did not come through the appropriate predecessor in-port,
the token is ignored. 
\end{itemize}
Now, is the resulting structure a ring-of-trees? Unfortunately, not
yet. For processors $x$ and $y$, let $d'(x,y)$ be the distance
from $x$ to $y$ within the structure; let $L$ be the length of
the marked ring. (Recall that we assume that $\Gamma <L$.) For the
structure to be a ring-of-trees, we need $d'(root,y)<L$ for all leaves
$y$ in the network. This is not necessarily true. However, we can
prove:

\begin{lemma}For any leaf processor $y$, $d'(root,y)<2L$\end{lemma}

\emph{Proof}. Fix any leaf processor $y$ and any ring processor $r$.
If there exists a shortest path from $r$ to $y$ that does not intersect
the ring (other than at $r$), then by construction\[
d'(root,y)\le d'(root,r)+d(r,y)\]
Consider a shortest path from $root$ to $y$ (not necessarily within
the structure). Such a path must have length less than or equal to
$\Gamma $. Let the final ring node that this shortest path intersects
be $final$. By the above observation, we have\[
d'(root,y)\le d'(root,final)+d(final,y)\]
If $final=root$, then the expression is less than or equal to $\Gamma $
and hence less than $2L$. If $final\neq root$, then $d'(root,final)\le L$
and \[
d(final,y)\le d(root,y)-1<\Gamma \le L\Rightarrow \]
\[
d'(root,y)\le d'(root,final)+d(final,y)<2L\]
\qed

How can we use this fact? First, we note that once a token has traveled
around the ring twice, the root can rest assured that the TREE-PARENT
tokens have had opportunity to reach every processor in the network;
hence every processor has a parent designation, as required. Second,
we can use this fact to create a ring-of-trees from the above structure.

To do so, we \char`\"{}split\char`\"{} each of the ring processors
into two distinct personalities -- say, personalities \char`\"{}0\char`\"{}
and \char`\"{}1\char`\"{}. The root accomplishes this by sending a
SPLIT token around the loop simultaneously with the TREE-PARENT token.
All messages from processors on the loop have the number of the personality
sending it attached. If a non-root processor $n$ receives a message
from personality $i$ of its neighbor, then personality $i$ of processor
$n$ handles it; if the root receives a message from personality $i$
of its neighbor, then personality $i+1$ mod 2 handles it. If we decree
that every \char`\"{}tree\char`\"{} hanging from a ring processor
in our structure hangs from personality 0, then we can simulate a
twofold increase in the length of our marked loop, which is all we
need for a ring-of-trees.

The astute reader will note that there is nothing special about the
number \char`\"{}2\char`\"{} in the above argument; any fixed constant
$k$ would do in its stead. We thus have

\begin{lemma}\label{lemma:split}If we can mark a loop of length
$\Theta (\Gamma )$, then we can fire the whole network; indeed, once
we have such a marked loop, it takes only time $O(\Gamma )$ to fire
it.\end{lemma} 

\emph{Proof.} Given a marked loop of size $\Theta (\Gamma )$, we
can split it in time $O(\Gamma )$, form a ring-of-trees structure
in time $O(\Gamma )$, then fire it (using the protocol in \cite{HN})
in time $O(\Gamma )$.\qed

It remains only to show that we can construct such a marked loop quickly;
this will be the main thrust of our protocol (see Section \ref{section:algorithm}).

\section{Overview of the Protocol\label{section:overview}}

In this section, we assume that we are given a Wake Up and Report
Protocol that runs in time $W$, where $W$ is dependent on the network
topology and size. From this protocol, we will construct a Firing
Squad Synchronization Protocol that runs in time $CW$, where $C$
is a constant \emph{independent} of network topology and size. Because
a protocol for the FSSP is necessarily a protocol for the WURP%
\footnote{Clearly, a processor must be non-quiescent in order to fire. Hence,
upon firing, the root can be certain that there are no quiescent processors
in the network.%
}, this will be enough to prove the asymptotic time-equivalence of
the two problems. 

Before proceeding to the detailed steps of the constructed protocol
for the FSSP, we will present a top-down version which illustrates
the basic ideas. The given protocol for the WURP is used as a stopwatch.
Since, by lemma \ref{lemma:LBL}, its running time $W$ is $\Omega (\max (\Gamma ,\Gamma '))$,
we know that sufficiently many repetitions of the WURP will provide
enough time for a message to travel from the root to any given processor
(or vice-versa); this will be crucial in establishing the running
time of our protocol. 

While repeating the WURP, we build and dismantle increasingly longer
loops through the root; each successive loop will be at least 4 times
the length of its predecessor. After sufficiently many iterations
of the WURP -- i.e., in time $O(W)$ -- we are guaranteed (for reasons
to be explained in Section \ref{section:proofofcorrectness}) that
the length of the current loop is at least $\frac{\Gamma }{4}$. By
lemma \ref{lemma:split}, once we have such a loop, we can fire it
in time $O(\Gamma )$; since $W$ is $\Omega (\Gamma )$, this suffices
to prove the main result of this paper. 

We will now elaborate on some of the more important steps in the brief
description above. The process of constructing and dismantling increasingly
longer loops in the network is accomplished via the use of various
types of snakes. Growing snakes are sent out, and return carrying
the encoded paths that allow the construction of the loops. (The steps
for this are outlined more fully below.) When construction of a new
loop begins, the previous loop is dismantled; when construction completes,
we wipe out all incipient loops within a certain radius of the root,
thus guaranteeing that the next loop, if any, will have length at
least 4 times that of the current one. When the WURP {}``stopwatch''
calls time, we finish whatever loop we are working on, if any, assured
that the last loop we mark will be extremely long; indeed, as mentioned
above, its length will be at least $\frac{\Gamma }{4}$.

\section{The Protocol (detailed description)\label{section:algorithm}}

Recall that we are given a protocol for the Wake Up and Report Problem.
From this protocol, we will construct a protocol for the Firing Squad
Synchronization Problem which runs in no more than a constant multiple
of the time-complexity of the given WURP protocol. 

\begin{enumerate}
\item The root processor is signaled to commence the FSSP protocol. At this
time it does two things simultaneously. First, the root commences
the given WURP protocol for the first of {}``sufficiently many''
times%
\footnote{For those who crave explicitness, we will show later that {}``sufficiently
many'' is no more than 8. It may well be possible to reduce this
number by making the following protocol more efficient; an improvement
of this sort will not affect our final asymptotic result, however.%
} (when it finishes, the root immediately starts it again). Simultaneously,
the root begins the protocol described in steps \ref{item:first}
through \ref{item:last} below%
\footnote{We assume that the two protocols do not interfere with each other;
running two protocols at once merely requires us to enlarge the language
and state space by a constant factor, since each requires only a constant
amount of language and state space.%
}.\\
\\
 \noun{Steps \ref{item:first} through \ref{item:endloop} (except
\ref{item:both}) will be devoted to marking a loop through the network
that includes the root. Step \ref{item:both} is devoted to clearing
the extraneous snake markings created during this process.}
\item \label{item:first}The root releases out-growing snakes. Any processor
receiving an out-growing snake character for the first time marks
itself as {}``OG-visited'', thus preventing any subsequent OG-snakes
from entering it. It will also designate the in-port from which it
received the OG-snake as its {}``OG-parent'' in-port. The {}``OG-visited''
mark will never be cleared; hence, the OG-snakes carve out a breadth-first-search
tree.\\
\\

\item \label{item:IG-start} A processor that receives an OG-snake will
also create a corresponding in-growing snake, unless an already-created
IG-snake accompanies the OG-snake, in which case the processor will
pass the latter along, in accordance with growing-snake rules. In
either case, any OG-snake character broadcast by the processor will
be accompanied by the corresponding IG-snake character; e.g., if the
processor broadcasts the character OG(2), it also must broadcast the
character IG(2).

In-growing snakes do not respect the OG-visited and OG-parent markings
created by OG-snakes; instead, they leave analogous {}``IG-visited''
and {}``IG-parent'' markings. In addition, any node that creates
an IG-snake marks itself {}``IG-start.'' Unlike the OG-markings,
however, these IG-markings may later be deleted, along with the snakes
that made them (see step \ref{item:both}). 

We note for clarity that, once created, an IG-snake head will accompany
any given OG-snake head wherever the latter goes, until such time
as the IG-snake is deleted. Thus, until that deletion occurs, no new
IG-snakes will be created. For example, when the protocol first begins,
IG-snakes will be created at all processors whose distance from the
root is one, and nowhere else until the IG-snakes are deleted. Once
this deletion occurs, new IG-snakes will be created en masse if and
when the OG-snakes reach as-yet-unvisited processors. Until its deletion
begins, a generation of IG-snakes will carve out a forest of breadth-first-search
trees. 

\item \label{item:mark} When the root receives the head of an IG-snake,
it does three things. First, it closes itself off to all other IG-snakes,
ignoring any that attempt to enter (this closure will continue until
step \ref{item:endloop}, even if the root's IG-visited mark is cleared
in the interim). Second, if a loop has already been marked within
the network%
\footnote{If no loop has yet been marked, ignore this step.%
}, the root begins the process of unmarking it; it does so by sending
the speed-1 loop token UNMARK around the loop. Each processor in the
old marked loop, upon receiving the token through its appropriate
predecessor in-port, passes the token through the appropriate successor
out-port, then forgets those predecessor and successor designations%
\footnote{We note here that, although the unmarking of the old loop will be
complete before we finish marking the new loop, the loops may still
interfere. However, we will never have more than one loop in the network
at a time. Each processor must be able to keep track of two loops,
locally (i.e., which edges are incident to each loop.) The root (which
is in every loop) must keep track of a parity bit to distinguish between
loops; it should send this out with every dying-snake character or
loop token so that processors on the loop can tell which loop it affects.
This is an annoying but unimportant complication, and we will say
no more about it.%
}. Third, we note that the characters of the incoming IG-snake encode
a path from the root back to itself; this will be our new loop. The
root then converts this IG-snake into the corresponding OD-snake (i.e.,
the root must therefore eat the head character of the IG-snake as
if it were an OD-snake character, then send the rest of the snake
through the appropriate out-port, converting each IG-snake character
into the corresponding OD-snake character.) The OD-snake will mark
that part of the loop generated by the OG-snake; when the OD-snake
reaches a processor marked IG-start, the process converts it an ID-snake,
which proceeds to mark that part of the loop generated by the IG-snake.%
\footnote{We convert OD-snakes to ID-snakes so as to avoid complications when
that part of the loop generated by the OG-snake intersects the part
generated by the IG-snake; see Sections \ref{subsubsection:dyingsnakes}
and \ref{subsection:markedloops} for loop-marking details. Notice
that the processor marked IG-start cannot appear twice on the loop,
since it ends the first part and starts the second, and neither part
is self-intersecting; this justifies an assertion made in Section
\ref{subsubsection:dyingsnakes}.%
} The ID-snake will die as it returns to the root; that is, the root
will receive just the tail of the ID-snake. 
\item \label{item:both} When the marking of the new loop is complete (this
occurs when the tail of the ID-snake reaches the root), the root does
two things. First, it sends the speed-1 loop token CLOCK around the
new loop. This token does nothing except get passed from processor
to processor along the marked loop until it returns to the root. Second,
we now want to remove the IG-snake characters and markings still percolating
through the network. To this end, the root broadcasts breadth-first
speed-4 PREPARE-TO-KILL tokens that propagate along the OG-tree. (Any
processor receiving a PREPARE-TO-KILL token that does not come from
its OG-parent ignores the token.) The PREPARE-TO-KILL tokens will
simultaneously reach the processors that have an IG-start marking.
(This will be proven later.) When this occurs, they change into breadth-first
speed-4 KILL tokens that propagate along the IG-tree, deleting IG-snake
characters and clearing IG-start, IG-visited, and IG-parent markings
as they go%
\footnote{This seems to be a great deal of effort, and one might wonder why
such effort is justified. Note that if we are not on our first marked
loop, then there must have been at least one generation of IG-snakes
that has been deleted by KILL tokens. Many processors -- including,
perhaps those near the out-ports of the root -- may have had their
IG-visited markings cleared at that time and never replaced by a subsequent
IG-snake. If we simply allow the root to release KILL tokens, they
may run into these non-IG-visited processors and be ignored. Thus
we arrange matters so the KILL tokens will be activated at those processors
with IG-start markings; these processors serve as the bases for the
trees in the IG breadth-first forest, so are a convenient place to
start the deletion of IG-snakes and associated markings.%
}. 
\item \label{item:endloop} The CLOCK token will return to the root at the
same moment that the KILL tokens finish deleting all IG-snakes and
markings from the network. At this point, the root opens itself up
to IG-snakes once more. Let $L$ be the length of the just-completed
loop. If there are still quiescent processors in the network, then
we make two claims: first, the distance from the root to the most
distant non-quiescent processor is exactly $4L$; second, a new generation
of IG-snakes will be simultaneously spawned when the OG-snakes subsequently
reach distance $4L+1$ from the root. For a proof of these facts,
see Lemma \ref{lemma:loopinvariants}; for a visualization of this
step, see Figure \ref{figure:algorithm}.%
\begin{figure}
\includegraphics[  scale=0.5]{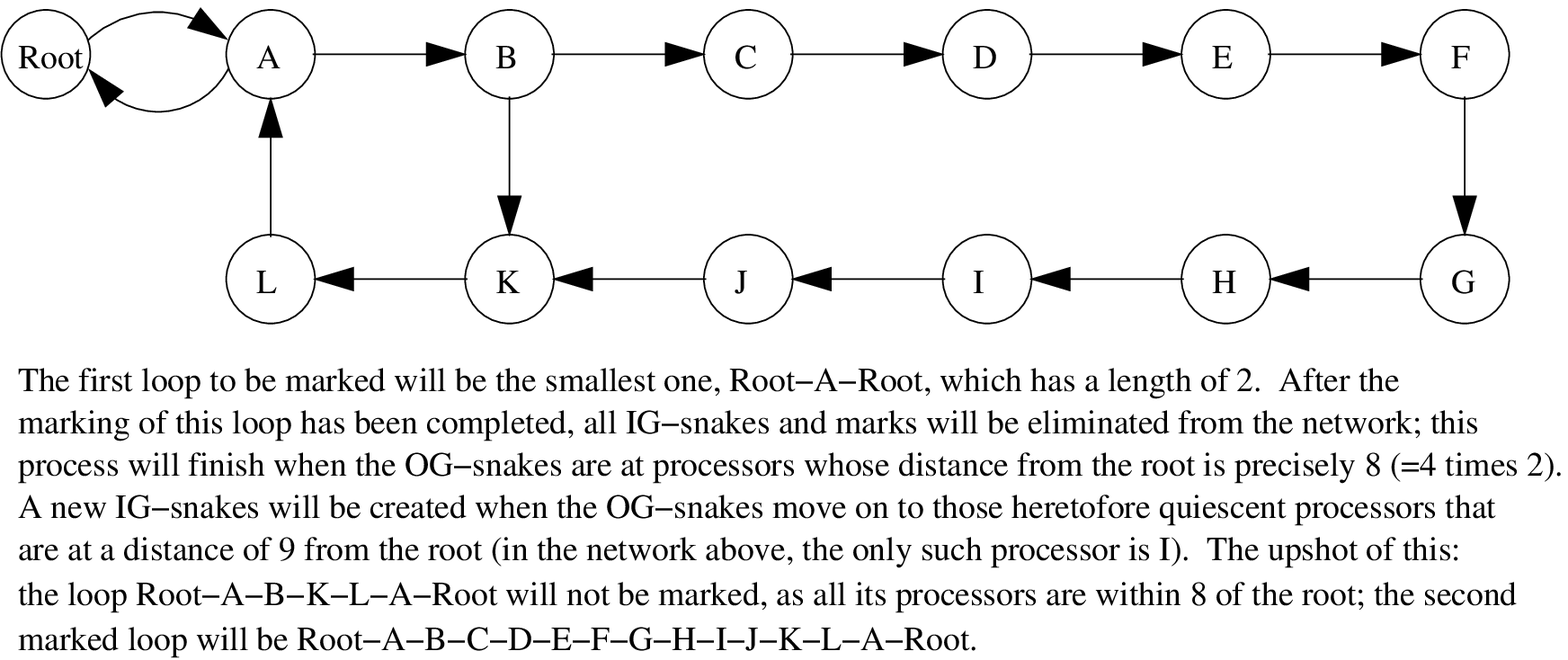}

\caption{\label{figure:algorithm}The main idea of step \ref{item:endloop}}
\end{figure}

\item \label{item:finalmark}The root repeats steps \ref{item:mark} through
\ref{item:endloop} until the given WURP protocol has run sufficiently
many times (where what {}``sufficiently many'' is will be determined
later). At that time, the root completes whatever loop, if any, it
is working on, and eliminates all IG-snakes and markings remaining
in the network. 
\item \label{item:last}At this point, we have a single marked loop in the
network. This loop, we will show, must have length at least $\frac{\Gamma }{4}$.
As described in Section 3.5, we can now split this loop fivefold,
create a ring-of-trees from the lengthened loop, and then fire the
ring-of-trees. 
\end{enumerate}

\section{Proof of the Protocol's Correctness\label{section:proofofcorrectness}}

\subsection{Lemmas. }

\begin{lemma}\label{lemma:loopinvariants}

Let \(L\) be the length of the current marked loop (if no loop has been marked
yet, let \(L\) be zero.)  We claim the following loop invariants:

\begin{enumerate}

\item Let \(T\) be the number of time ticks elapsed since the start of our
protocol.  At \(T = 3(4L)\), the marking of this loop was completed.
At \(T = 4(4L)\), all IG-snakes were
eliminated from the network, along with all IG-visited,
IG-start, and IG-parent markings; one time 
tick later, all KILL tokens were gone.

\item The next generation of IG-snakes (if any) will all be created 
at \(T = 4(4L + 1)\), at all processors whose distance from the root is
\((4L + 1)\); these processors will all be marked IG-start, of course.

\item The next loop (if any) will have length greater than \(4L\).

\end{enumerate}

\end{lemma}

\begin{proof}
We proceed by induction on the number of loops.  All claims are trivial
when \(L=0\).  Now, how much time had elapsed when we finished marking the
current loop?  Well, \(4L\) had passed when the head of the IG-snake that 
encoded the loop reached the root\footnote{The ubiquitous 4's in these calculations (well, most of them)
are because snakes travel at speed 1.};
\(4(3L)\) had passed by the time the tail of the ID-snake reached the root, 
indicating the end of the loop-marking process.  At this point, the speed-4
PREPARE-TO-KILL were released; we may assume (by the inductive hypothesis)
that they simultaneously became speed-4 KILL tokens when they reached the
processors marked IG-start.  These KILL tokens caught up to the
IG-snake heads farthest from the root at \(T=4(4L)\).  Were they able to
eliminate all IG-snakes and markings
en route?  Certainly; as noted in Section \ref{section:algorithm}, before
the deletion process begins, a 
generation of IG-snakes carves out a forest of breadth-first-search
trees anchored at those processors marked IG-start.  Our KILL tokens began
working at those processors and raced their way up the trees, annihilating
all IG-snakes and markings\footnote{It's possible for an IG-snake to ``double-back'' to 
previously-visited processors once the deletion process has begun; nonetheless,
the region between a KILL token and the heads of its associated IG-snakes
will always be a tree, so the KILL token will have no difficulty eliminating
all IG-snakes and markings, even those snakes that double-back.  We point out
that this lack of difficulty is a large advantage of activating the KILL
tokens at those processors marked IG-start.}.
One time tick after \(T=4(4L)\), before the OG-snake heads could move on, 
each of the KILL tokens fell off its associated IG-snake breadth-first-search
tree, so died.  

Now, at \(T=4(4L+1)\), the OG-snake heads move on, reaching all processors that
are a distance of \(4L+1\) from the root.  Since the OG-snake heads are 
unaccompanied by IG-snake heads, the processors simultaneously make a new
generation IG-snakes (and mark themselves IG-start.)  
As pointed out in step \ref{item:IG-start} of the 
protocol in Section \ref{section:algorithm}, no more will be created until the
generation is eliminated.

Finally, the first IG-snake from this generation that reaches the root will
form the next loop; this loop will be the shortest one in the network that
contains a processor that is at least \(4L+1\) from the root.  In any case,
however, the length of the loop will clearly be more than \(4L\).

\end{proof}

\begin{lemma}\label{lemma:existloop}

The root will receive at least one IG-snake (and hence start
the creation of a marked loop) within time \(4(\Gamma '+1)\) of the
beginning of the protocol.

\end{lemma}

\begin{proof}First, recall that all snakes are speed-1 and thus move
along edges every 4 clock ticks. After 4 clock steps, at least one
IG-snake -- call it \(s\) -- is released. Now, either snake \(s\) makes
it back to the root in at most \(4\Gamma '\) more clock ticks, or it is killed
along the way. However, since KILL tokens are only released from
the root after the root has received an IG-snake, the only way IG-snake
\(s\) can be killed is by running into an IG-snake
that is no farther away from the root than itself. But since a collision
of two snakes always results in the survival of one, at least one
IG-snake among those no farther away from the root than \(s\) must
survive to reach the root.
\end{proof}

\begin{lemma}\label{lemma:nomore}

The "sufficient number" of WURP iterations mentioned in Step 7 of our protocol is 8. After WURP has been run eight times, the root has completed whatever loop (if any) it is working on, and all IG-snakes and markings have been eliminated from the network, we can rest assured of two things: first, at least one marked loop will have been created; second, the OG-snakes will have visited every processor in the network.

\end{lemma}

\begin{proof} By Lemma 2.1, any WURP protocol takes time greater than $\max(\Gamma, \Gamma')$; thus 8 iterations must take at least time $T = 4(\Gamma+\Gamma')$. Note that $T \geq 4(\Gamma' + 1)$; thus, by Lemma 6.2, at least one marked loop will have been created. Note also that $T \geq 4(\Gamma + 1)$, so we know that the OG-snakes will have visited every processor in the network.
\end{proof}

When step \ref{item:finalmark} is completed, we will have a unique
last marked loop in the network. Let $M$ be its length. 

\begin{lemma}\label{lemma:maintheorem}

\(4M\geq \Gamma \)

\end{lemma}

\begin{proof}Let \(T\) be the number of time ticks since the start
of the protocol.
When the last loop is completed and all the IG-snakes and markings are 
cleaned up, we have \(T/4=4M\) (by the 
same logic as in the proof of Lemma \ref{lemma:loopinvariants}).

We must also have $T/4 \geq \Gamma$ at this point, or else there would be non-OG-visited processors in the network, which would contradict Lemma 6.3. We therefore have $4M=T/4 \geq \Gamma$, which is the result we desire.
\end{proof}

So we have a loop of length at least $\frac{\Gamma }{4}$, which is
$\Omega (\Gamma )$; we are done.

\subsection{Proof of the Protocol's Running Time. }

\begin{theorem}

Given any WURP protocol (with running time \(W\)), the protocol presented
in Section \ref{section:algorithm} creates a FSSP protocol which
runs in time \(O(W)\).

\end{theorem}

\begin{proof} Fix a network and WURP protocol; let \(W\) be its running
time.  We proved in Lemma \ref{lemma:LBL} that \(W\) is 
\(\Omega(\max(\Gamma ,\Gamma '))\), so any \(O(\Gamma )\) process is certainly
\(O(W)\).
The main loop of the protocol
takes \(8W\) time (recall that we found 8 iterations of the WURP sufficient); 
marking the final loop is certainly an \(O(W)\)
process, as is splitting it; finally, firing the network takes \(O(W)\)
time. Thus, the protocol takes \(O(W)\) time, as claimed. \end{proof}

\section{Asymptotic Time-Equivalence of Other Problems}

We have shown that the Firing Squad Synchronization Problem and the
Wake Up and Report Problem are asymptotically time-equivalent. Other
common network protocols can be easily shown to be asymptotically
running-time equivalent to these. In this section, we illustrate several
immediate examples. We specify that in each problem below, the root
must enter a specific terminal state before the protocol can be considered
completed. Thus, it is implicit in all of the following protocols
that the root must be informed when the network has completed its
assigned task before the protocol terminates.

\subsection{Build a Rooted Outgoing Spanning Tree. }

The tree created by the out-growing snakes in our Firing Squad Synchronization
Protocol can be utilized as a rooted outgoing spanning tree. Conversely,
if all processors must have parent designations, they must have awakened
at some point to get their designations; thus the protocol for building
a rooted spanning tree can also be used as a Wake Up and Report Protocol.

\subsection{Long Circuit/Slow Clock. }

The Long Circuit/Slow Clock problem asks for a rooted outgoing spanning
tree and a loop through the network such that the length of the loop
is at least a constant multiple of the length of the longest branch
of the tree. The loop created in our Firing Squad Synchronization
Protocol can be used as a Long Circuit/Slow Clock; conversely, given
a Long Circuit/Slow Clock, one can easily construct a Wake Up and
Report Protocol (simply send a token sufficiently slowly around the
circuit while broadcasting a wake-up message so that one can be sure
that every processor has awakened by the time the token returns.)

\subsection{Conducting a Network Search. }

The objective of conducting a Network Search is for the root to query
the network for some piece of {}``information.'' This information
can be stored in the form of bits in one or more processors. The root
terminates the protocol when either (a) it gets notification that
the information it requires exists in some specific processor in the
network and creates a marked path to said processor or (b) it finds
that the information it requires does not exist anywhere in the network. 

One possible protocol which solves the Network Search Problem is as
follows: Have the root broadcast a breadth-first query token to the
network. Simultaneously, have the root initiate a WURP protocol. The
protocol for each of the processors in the network that contain the
required bits of information will be to simply broadcast IG-snakes
which follow exactly the same rules as outlined in Section \ref{section:algorithm}
above. Otherwise, simply propagate the root's query message. 

If the root does not receive the head of a IG-snake by the time 8
WURP protocols have been executed, the root can terminate the protocol
with the knowledge that the information it requires does not exist
anywhere in the network. We know that any WURP protocol must take
longer than $\max (\Gamma ,\Gamma ')$. Thus 8 executions must take
time greater than $4(\Gamma +\Gamma ')$, which is an upper bound
for the amount of time that it would take for the query to reach any
given processor and have any potential IG-snake return. 

If the root does happen to receive the head of a IG-snake before 8
WURP protocols have been executed, it simply establishes the marked
path as illustrated in Section \ref{section:algorithm}. 

In either case, the running time is at most a constant multiple of
the given WURP protocol's running time. 

Clearly, if the information is not to be found in the network, a Network
Search must query every processor. Thus, at termination, every processor
must have been awakened. We have therefore established the required
equivalence.

\section*{Acknowledgments}

The authors would like to thank the various referees, who pointed
out several protocol simplifications and presentation suggestions.

\bibliographystyle{plain}
\bibliography{equivalent}

\end{document}